\documentclass[aps,prl,floatfix,showpacs,twocolumn,amsmath,amssymb,superscriptaddress,nofootinbib]{revtex4-2}

\usepackage{upgreek}
\usepackage{hyperref}
\usepackage{siunitx}
\usepackage{amsmath}
\usepackage{graphicx}
\usepackage{float} 

\begin{document}

\title{Observation of Rayleigh optical activity for chiral molecules: a new chiroptical tool}

\author{Duncan~McArthur}
\email{These authors contributed equally.}
\affiliation{SUPA and Department of Physics, University of Strathclyde, Glasgow, G4 0NG, UK}
\author{Emmanouil~I.~Alexakis}
\email{These authors contributed equally.}
\affiliation{SUPA and Department of Physics, University of Strathclyde, Glasgow, G4 0NG, UK}
\author{Andrew~R.~Puente}
\affiliation{Department of Chemistry, Vanderbilt University, Nashville, TN 37235, USA}
\author{Rebecca~McGonigle} 
\affiliation{Department of Pure and Applied Chemistry, University of Strathclyde, Glasgow, G1 1RD, UK}
\author{Andrew~J.~Love}
\affiliation{James Hutton Institute, Invergowrie, Dundee DD2 5DA, UK}
\author{Prasad~L.~Polavarapu}
\affiliation{Department of Chemistry, Vanderbilt University, Nashville, TN 37235, USA}
\author{Laurence~D.~Barron}
\affiliation{Department of Chemistry, University of Glasgow, Glasgow G12 8QQ, UK}
\author{Lewis~E.~MacKenzie}
\affiliation{Department of Pure and Applied Chemistry, University of Strathclyde, Glasgow, G1 1RD, UK}
\author{Aidan~S.~Arnold}
\affiliation{SUPA and Department of Physics, University of Strathclyde, Glasgow, G4 0NG, UK}
\author{Robert~P.~Cameron}
\email{robert.p.cameron@strath.ac.uk}
\affiliation{SUPA and Department of Physics, University of Strathclyde, Glasgow, G4 0NG, UK}

\begin{abstract}
By measuring a small circularly polarized component in the scattered light, we report the first observation of Rayleigh optical activity (RayOA) for isotropic samples of chiral molecules, namely the two enantiomers of $\alpha$-pinene in neat liquid form. Our work validates fundamental theoretical predictions made over fifty years ago and expands the chiroptical toolkit.
\end{abstract}

\maketitle

\section{Introduction}
It was foreseen over half a century ago that an isotropic sample of chiral molecules should exhibit Rayleigh optical activity (RayOA) in the form of a small difference in the intensity of Rayleigh scattering in right- and left-circularly polarized incident light, now called incident circular polarization (ICP) RayOA, or, equivalently, a small circularly polarized component in the Rayleigh scattered light, now called scattered circular polarization (SCP) RayOA \cite{Atkins69a, Barron71a}. This followed from the theoretical discovery of a new light scattering mechanism from chiral molecules involving interference between light waves scattered via the polarizability and optical activity tensors \cite{Atkins69a}. The Raman optical activity equivalent (ROA) was observed quite quickly \cite{Barron73a} and has evolved into a powerful probe of the stereochemistry of chiral molecules and the structure and behaviour of biomolecules \cite{Barron15a}; but despite possible applications ranging from the robust assignment of absolute configuration \cite{Zuber08a} to the remote sensing of bioaerosols \cite{Gasso22a, Pan24a}, RayOA has remained largely overlooked. Measurements of optical activity in quasi-elastic light scattering have been published for a handful of biological macromolecules and structures \cite{Maestre82a, Pan24a}, but observation of RayOA has not been reported before for isotropic samples of chiral molecules like $\alpha$-pinene in neat liquid form, as considered here \cite{Barron04a}.

\section{Method}
We constructed a dedicated RayOA instrument to remedy this remarkable omission and facilitate the full exploitation of RayOA as a chiroptical tool. In theory, RayOA is simpler than other chiroptical methods like ROA \cite{Nafie11a, Lightner24a, Er24a} and circularly polarized luminescence (CPL) \cite{Stachelek22a, Baguenard23a} in that the Rayleigh scattered light has essentially the same wavelength as the incident light. In practice, this quasi-degeneracy makes it difficult to distinguish RayOA from stray light artifacts, necessitating a bespoke instrumental design. Our instrument employs the SCP strategy using an incident narrow-linewidth laser beam with wavelength $\lambda=532\,\mathrm{nm}$, linearly polarized in the scattering plane, propagating through an isotropic sample of chiral molecules held in a cuvette. A high-precision detection system collects a small solid angle of $2\times 10^{-5}\,\mathrm{sr}$ of the Rayleigh scattered light at right-angles and measures the SCP observable $\Delta$ as the difference in the intensity of the right- and left-circularly polarized components divided by their sum; $\Delta$ should have equal magnitudes but opposite signs for enantiomers. This is equivalent to the depolarized ICP strategy with the right-angle scattered light collected through a linear polarizer parallel to the scattering plane, as used in the original ROA observations \cite{Barron73a, Barron15a}. In both cases this suppresses the isotropic scattering that is responsible for large polarization artefacts. For an enantiopure neat sample of conformationally rigid chiral molecules illuminated far off resonance, well-established theory predicts that \cite{Barron04a} 
\begin{align}
\Delta\approx{}&\frac{1}{c}\frac{24\beta(G^\prime)^2-8\beta(A)^2}{12\beta^2},
\label{Delta}
\end{align}
where $c$ is the speed of light and $\beta(G^\prime)^2$, $\beta(A)^2$, and $\beta^2$ are rotational invariants that quantify the anisotropy of each molecule, with the polarizability/electric dipole-magnetic dipole optical activity anisotropy $\beta(G^\prime)^2$ and the polarizability/electric dipole-electric quadrupole optical activity anisotropy $\beta(A)^2$ being chirally sensitive \cite{Barron04a, Zuber08a}. This dimensionless RayOA observable enables us to avoid the theoretical challenge of describing isotropic Rayleigh scattering in dense samples \cite{Cameron24a} and can be regarded as a measure of pure rotational (as opposed to vibrational) ROA in that the numerator and denominator of eq \ref{Delta} are integrated rotational Raman difference- and sum-frequency spectra, respectively \cite{Young82a, Barron85a, Polavarapu87a}. According to eq \ref{Delta}, $\Delta$ varies simply with wavelength far off resonance as $\Delta\propto1/\lambda$ to good approximation. Measurements of $\Delta$ at a single wavelength ($\lambda=532\,\mathrm{nm}$ here) are therefore sufficient to extract the chiroptical information available.

The optimized geometry of $\alpha$-pinene used for calculations here was taken from the OR45 benchmark performed by Autschbach and coworkers \cite{Srebro11a}, which was optimized at the B3LYP/6-311G(d,p) level. The three polarizability tensors (electric dipole-electric dipole, electric dipole-magnetic dipole, and electric dipole-electric quadrupole) needed for predicting the rotational invariants $\beta(G^\prime)^2$, $\beta(A)^2$, and $\beta^2$ and thus the circular intensity differential $\Delta$ were computed using the Gaussian 16 program \cite{Gaussian16a}. More details on the implementation of RayOA calculations are given in reference \cite{Puente24a}. Calculations for $\alpha$-pinene were done in the gas-phase with several popular density functionals (CAM-B3LYP \cite{Yanai04a}, $\omega$B97X-D \cite{Chai08a}, LC-wHPBE \cite{Henderson09a}, B3LYP \cite{Lee88a, Becke93a}, M06-2X \cite{Zhao08a}, and B3PW91 \cite{Perdew92a}) utilizing the augmented correlation-consistent Dunning basis set, aug-cc-pVTZ \cite{Dunning89a, Kendall92a}. Calculations were performed with various discrete input wavelengths, which were chosen based on common laser setups and optical rotation (OR) measurements: $365$, $405$, $436$, $532$, $546$, $589$, $633$, $799.3$, and $1064\,\mathrm{nm}$.

Our investigations \cite{Puente24a} on a variety of molecules indicate that RayOA is fairly insensitive to the level of theory used, while predicted OR can vary strongly with the level of theory. This is in agreement with the conclusions of Zuber et al., who stated that RayOA has only weak dependence on the electronic structure method and basis set \cite{Zuber08a}. In addition, we found \cite{Puente24a} that RayOA is fairly insensitive to conformational flexibility, incremental explicit solvation, and dispersion interactions. These attributes make it much easier to determine the absolute configuration of chiral molecules using RayOA than other chiroptical methods, such as OR, circular dichroism (CD), and ROA.

\section{Results and Discussion}
\begin{figure}[!t]
\centering\includegraphics[width=\linewidth]{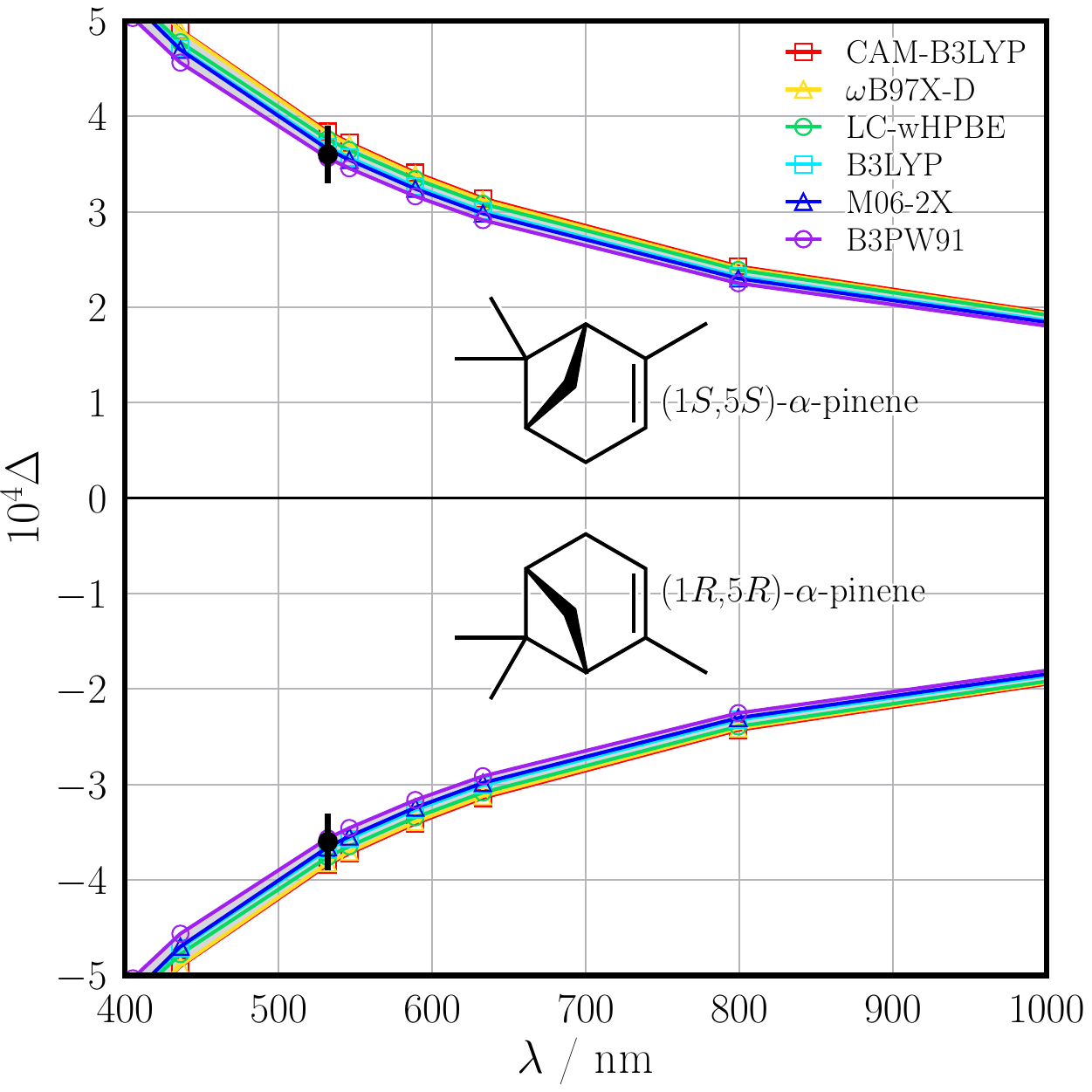}
\caption{Measured (black data points) and predicted (colored data points connected by straight lines) values of the circular intensity differential $\Delta$ for the enantiomers of $\alpha$-pinene.}
\label{Figure1}
\end{figure}

In Figure \ref{Figure1} we show our measured values of the circular intensity differential $\Delta$ for $\alpha$-pinene (analytical standard; $>99\%$ purity); $\Delta=(+3.6\pm0.3)\times 10^{-4}$ for the ($1S$,$5S$) enantiomer and $\Delta=(-3.6\pm0.3)\times 10^{-4}$ for the ($1R$,$5R$) enantiomer. These have been corrected for small enantiomeric imbalances in our samples, slight rotation of the incident light as it propagated through the samples, and a small enantiomer-independent offset in our instrument. Our measured values are in excellent agreement with our predicted values of eq \ref{Delta}, which range from $\Delta=\pm 3.56\times 10^{-4}$ to $\Delta=\pm 3.84\times 10^{-4}$ with an average of $\Delta=\pm 3.72\times 10^{-4}$ across our chosen computational methods; we attribute the small possible discrepancy in magnitude to impurities. Importantly, our results demonstrate that the measured signs of $\Delta$ at a single wavelength of $\lambda=532\,\mathrm{nm}$ are sufficient to correctly determine the absolute configuration of $\alpha$-pinene, the predicted signs being robust against choice of computational method \cite{Zuber08a}. Our measured values were distilled from data accumulated during total exposure times of $15.0\,\mathrm{hr}$ ($1S$,$5S$) and $18.3\,\mathrm{hr}$ ($1R$,$5R$) with laser powers of $30\,\mathrm{mW}$ and $1/e^2$ beam diameters of $0.2\,\mathrm{mm}$ incident upon the cuvettes to approach the precision limit ($\sim 10^{-5}$) of our instrument. These exposure times can be significantly reduced using higher laser powers and/or by collecting larger solid angles of Rayleigh scattered light.

\section{Summary and Outlook}
By employing the SCP strategy, we have reported the first observation of RayOA for isotropic samples of chiral molecules. Our work expands the chiroptical toolkit to include RayOA as a simple new method for the determination of absolute configuration that is more reliable than OR \cite{Zuber08a, Nafie11a}, and easier than other chiroptical methods \cite{Puente24a}. It is noteworthy that RayOA offers chirally sensitive electric quadrupole information ($\beta(A)^2$) even for isotropic samples; the extraction of such information using OR or CD instead requires molecular orientation \cite{Barron04a}.

RayOA is not to be confused with \textit{hyper} Rayleigh scattering optical activity (HRS OA), which was also first predicted decades ago \cite{Andrews79a} but has only been observed recently \cite{Collins19a, Ohnoutek21a}. 

The backscattering in-phase dual circular polarization (DCP$_\mathrm{I}$) strategy has proven particularly advantageous for the measurement of ROA \cite{Nafie89a, Che91a, Nafie11a}. A backscattering DCP$_\mathrm{I}$ rather than right-angle SCP strategy might prove similarly useful for the measurement of RayOA, however we note that in a backscattering geometry there is the added challenge of distinguishing the Rayleigh scattered light of interest (due to sample fluctuations \cite{Cameron24a}) from incident light that has been retroreflected at air-cuvette and cuvette-sample interfaces as well as Rayleigh scattered light from within the walls of the cuvette, as all have essentially the same wavelength. In contrast, the Raman scattered light of interest in backscattering ROA can be isolated simply using a notch filter centred on the wavelength of the incident light.

The precise design of our instrument together with additional measurements will be reported elsewhere in due course.

\section{Author Contributions}
\textbf{D.~McArthur:} Instrument design and construction, data acquisition for $\alpha$-pinene. \textbf{E.~I.~Alexakis:} Instrument design and construction, data analysis, first observation. \textbf{A.~R.~Puente:} Quantum-chemical calculations. \textbf{R.~McGonigle:} Sample preparation. \textbf{A.~J.~Love:} Sample preparation. \textbf{P.~L.~Polavarapu:} Quantum-chemical calculations. \textbf{L.~D.~Barron:} Conceptualization, theory. \textbf{L.~E.~MacKenzie:} Sample preparation. \textbf{A.~S.~Arnold:} Instrument design and construction, data analysis, supervision. \textbf{R.~P.~Cameron:} Funding acquisition, instrument design and construction, supervision, original manuscript draft.

\section{Acknowledgements}

RPC gratefully acknowledges the support of a Royal Society University Research Fellowship  (URF$\backslash$R1$\backslash$191243, URF$\backslash$R$\backslash$241008, RF$\backslash$ERE$\backslash$210170, RF$\backslash$ERE$\backslash$231130) and EPSRC DTP (EP/R513349/1). LM gratefully acknowledges the support of a Royal Society University Research Fellowship (URF$\backslash$R1$\backslash$251567).The authors thank Carin Lightner, David McKee, Paul Griffin, and Anna Gribbon for helpful discussions.

\bibliography{References}

\end{document}